\documentclass[twocolumn,amsmath,prd,nofootinbib,superscriptaddress]{revtex4}
\usepackage{epsfig}
\usepackage{dsfont}
\usepackage{hyperref}

\def\beq{\begin{equation}}
\def\eeq{\end{equation}}
\def\bea{\begin{eqnarray}}
\def\eea{\end{eqnarray}}
\def\nn{\nonumber}

\def\tev{\rm TeV}
\def\gev{\rm GeV}

\def\fb{\rm fb}
\newcommand{\br}{{\tt Br}}
\newcommand{\til}{\widetilde}
\newcommand{\lsim}{\mathrel{\hbox{\rlap{\hbox{\lower4pt\hbox{$\sim$}}}\hbox{$<$}}}}
\newcommand{\gsim}{\mathrel{\hbox{\rlap{\hbox{\lower4pt\hbox{$\sim$}}}\hbox{$>$}}}}

\begin{document}

\title{Four-Lepton Resonance at the Large Hadron Collider}
\author{Vernon Barger}
\affiliation{Department of Physics, University of Wisconsin, Madison, WI 53706, USA}
\author{Hye-Sung Lee}
\affiliation{Department of Physics, Brookhaven National Laboratory, Upton, NY 11973, USA}
\date{November 2011}

\begin{abstract}
A spin-1 weakly interacting vector boson, $Z'$, is predicted by many new physics theories.
Searches at colliders for such a $Z'$ resonance typically focus on lepton-antilepton or top-antitop events.
Here we present a novel channel with a $Z'$ resonance that decays to 4 leptons, but not to 2 leptons, and discuss its possible discovery at the Large Hadron Collider.
This baryonic gauge boson is well motivated in a supersymmetry framework.
\end{abstract}

\maketitle

\section{INTRODUCTION}
Many models of physics beyond the Standard Model (SM) have an extra Abelian gauge group $U(1)$ \cite{Langacker:2008yv}.
There are many options for this $U(1)$ gauge symmetry and the corresponding $Z'$ from the broken symmetry can enable its identification.
The Drell-Yan process, wherein the $Z'$ is produced from quark-antiquark fusion and decays to a lepton-antilepton pair, can give a particularly clear signal at a hadron collider \cite{Collaboration:2011dca,CMS_dilepton}.  

However, the lepton pair search for a $Z'$ is nullified if the  $Z'$ does not couple to the SM leptons.
Searches can still be made for dijet decay products of a $Z'$, but the QCD dijet backgrounds are huge and fog such a signal \cite{preATLAS_dijet,preCMS_dijet}; hence, a $Z'$ resonance may not be discovered in dijets \cite{Chatrchyan:2011ns,Aad:2011fq}, especially if its coupling strength to quarks is not large, although a signal in the top pair channel could be easier to recognize \cite{CMS_ttbar,ATLAS_ttbar}.

\begin{figure}[b]
\begin{center}
\includegraphics[width=0.3\textwidth]{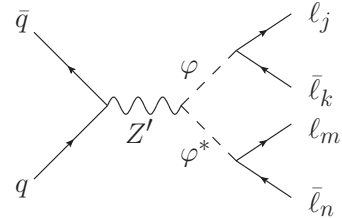} 
\end{center}
\caption{4-lepton $Z'$ resonance diagram at a hadron collider.}
\label{fig:4leptonResonance}
\end{figure}

Our interest here is in a 4-lepton signal from a leptophobic $Z'$ that can be produced at the LHC (and the Tevatron) with a large cross section and give a 4-lepton signal comparable to that of the lepton pair signals of generic $Z'$ models.
Specifically, we consider a $Z'$ resonance in which the 4-leptons final state is bridged by pair production of a new scalar boson ($\varphi$).
The $Z'$ couples to quark pairs and $\varphi$, but not to lepton pairs, and the new scalar $\varphi$ decays into a lepton pair (see Fig.~\ref{fig:4leptonResonance}).
LHC experiments, and possibly Tevatron experiments, can find or reject this distinctive 4-lepton signal.

A leptophobic $Z'$ may also appear as a resonance in a 6-lepton final state; a future search for this signal at the LHC requires $\sim 100 ~\fb^{-1}$ integrated luminosity at $14 ~\tev$ center-of-mass (CM) energy \cite{Barger:2009xg}.

\section{MODEL}
We begin by introducing a specific model in which a 4-lepton $Z'$ resonance can be realized without having a corresponding lepton pair signal.
We consider a generic supersymmetry (SUSY) framework where scalar fields are abundant.
The baryon number ($B$) is not preserved in the SUSY framework and in general the proton is unstable.
Thus a gauged $B$ has been sought, but then additional fermions are required to cancel the anomaly \cite{Carone:1994aa,Bailey:1994qv,FileviezPerez:2010gw,Ko:2010at,FileviezPerez:2011dg,Lebed:2011fw,FileviezPerez:2011pt}.
One natural way of anomaly cancellation is to add a fourth-generation (4G) of fermions.
Then, by requiring all quarks carry $B (=1/3)$, the 4G lepton charge is uniquely determined to be $-4$ by the anomaly free condition:
\beq
\begin{array}{rlcrc}
\text{SM quarks:}  & $1/3$, & ~ & \text{SM leptons:} & $0$,  \\
\text{4G quarks:}  & $1/3$, & ~ & \text{4G leptons:} & -$4$. \\
\end{array} \nn
\eeq
This is effectively $U(1)_B$ for the SM fermions: every SM quark has $B$ as a charge, and every SM lepton has $0$ charge\footnote{It was pointed out that the a baryonic $Z'$ can be a possible source of the $Wjj$ anomaly recently reported in the Tevatron CDF experiment \cite{Cheung:2011zt}.}.
	
Although proton stability would not have been guaranteed once the $U(1)_B$ is broken spontaneously, it turned out there exists a residual $\mathds Z_4$ discrete symmetry, called baryon tetrality ($\mathds B_4$), that forbids proton decay \cite{Lee:2010vj}.
Under $\mathds B_4$, lepton number violating operators can exist (such as $\lambda LLE^c$ and $\lambda' LQD^c$), but not baryon number violating operators (such as $\lambda'' U^cD^cD^c$).

In order to have the $\mathds B_4$ residual discrete symmetry, the Higgs boson that spontaneously breaks the $U(1)_B$ gauge symmetry (typically, a new Higgs singlet) should have a $U(1)_B$ charge of $4$ or $-4$ \cite{Lee:2010vj}.
Since it coincides with the $U(1)_B$ charge of the $N_4^c$ [4G right-handed neutrino and sneutrino (superpartner of neutrino)], we can adopt the approach of Ref.~\cite{Barger:2008wn} in which the 4G right-handed sneutrino (let us call it $S$) with a vacuum expectation value (vev) is used to break the $U(1)_B$ without the need for a separate singlet.
In general, the 4G sneutrino (left-handed one) can also have a vev through a mixing although we will assume the mixing is very small.

Because the 4G Majorana neutrino mass term is forbidden by the $U(1)_B$ symmetry, the 4G neutrino is a Dirac particle on which the seesaw mechanism does not work, and thus can easily satisfy the LEP $Z$ width measurement that is compatible only with 3 light active neutrinos \cite{Nakamura:2010zzi}.

We take the 4G sneutrino ($\til\nu_4$), the spin-0 companion of the 4G neutrino, as a bridging scalar between the $Z'$ and the lepton final states.
It has a nonzero $U(1)_B$ charge $(-4)$ and can couple to the $Z'$ while the sneutrinos of the first 3 generations have vanishing $U(1)_B$ charge.
We assume the $\til\nu_4$ is the Lightest Superpartner (LSP); it can decay into a lepton pair through the lepton number violating interaction $\lambda_{ijk} L_i L_j E^c_k$. (For instance, see Ref.~\cite{Barger:1989rk}.)
A nonzero $U(1)_B$ charge for the 3-generation sneutrinos would inevitably have led to 2-lepton $Z'$ resonance \cite{Lee:2008cn}.

In the remainder of this paper, we focus on the collider physics consequences of this scenario with the $\til\nu_4$ LSP.
Our analysis does not necessitate this particular supersymmetric model, albeit well motivated.
Rather, the model serves as an existence proof of a consistent theory of the 4-lepton $Z'$ resonance without a 2-lepton $Z'$ resonance, and also provides a specific realization of the phenomenology of a new scalar that couples to the $Z'$ and lepton pairs.

\section{LEPTONIC DECAY OF THE NEW SCALAR PARTICLE}
Here, we discuss some characteristic features of the $\til\nu_4$ LSP decay exclusively through lepton number violating operators.
Renormalizable operators $\lambda_{4jk} L_4 L_j E_k^c$ and $\lambda'_{4jk} L_4 Q_j D_k^c$ are forbidden by the $U(1)_B$ gauge symmetry.
Although operators with two 4G fields such as $\lambda_{4j4} L_4 L_j E^c_4$ and $\lambda'_{4j4} L_4 Q_j D^c_4$ (or $\lambda'_{44k} L_4 Q_4 D^c_k$) are allowed at the renormalizable level, $\til\nu_4$ decays cannot be mediated by these operators due to kinematics when $\til\nu_4$ is the lightest of the 4G states.
Thus, nonrenormalizable operators $\lambda_{4jk} \frac{\left<S\right>}{M} L_4 L_j E_k^c$ and $\lambda'_{4jk} \frac{\left<S\right>}{M} L_4 Q_j D_k^c$ with a heavy mass parameter $M$ allow $\til\nu_4$ decays since $z[S] = 4$.
Taking $\lambda_{4jk}$, $\lambda'_{4jk} \approx 1$ and $M / \left<S\right> = 10 - 1000$, for instance, effective coefficients $\lambda_{4jk}^{\rm eff} \equiv \lambda_{4jk} \frac{\left<S\right>}{M}$, ${\lambda'}_{4jk}^{\rm eff} \equiv \lambda'_{4jk} \frac{\left<S\right>}{M}\approx 0.001 - 0.1$ are obtained\footnote{$\left<S\right> \lsim 1~\tev$ is expected in SUSY to keep the extra $D$-term contribution to the sfermion masses small, and $M = 10 - 1000 ~\tev$ satisfies bounds on the scale of new physics from various constraints such as neutral kaon mixing.}.

Neglecting the light fermion masses, we obtain the partial widths
\bea
&&\Gamma(\til\nu_4 \to \ell^+_j \ell^-_k) = \frac{1}{16 \pi} (\lambda^{\rm eff}_{4jk})^2 m_{\til\nu_4} , \\
&&\Gamma(\til\nu_4 \to \bar d_j d_k) = \frac{3}{16 \pi} (\lambda'^{\rm eff}_{4jk})^2 m_{\til\nu_4} .
\eea
If we take all $\lambda' = 0$, the $\til\nu_4$ LSP would decay only through $\lambda_{4bc}^{\rm eff}$ ($b,c = 1 - 3$) with a total decay width given by
\bea
~~\Gamma_{\til\nu_4} = \frac{m_{\til\nu_4}}{16 \pi} \left[ (\lambda^{\rm eff}_{411})^2 + (\lambda^{\rm eff}_{412})^2 + (\lambda^{\rm eff}_{413})^2 + (\lambda^{\rm eff}_{421})^2 \right. \nn \\
~~\left. + (\lambda^{\rm eff}_{422})^2 + (\lambda^{\rm eff}_{423})^2 + (\lambda^{\rm eff}_{431})^2 + (\lambda^{\rm eff}_{432})^2 + (\lambda^{\rm eff}_{433})^2 \right] .
\eea
It is demanded that $m_{\til\nu_4} \gsim M_Z / 2$ by the result of the LEP $Z$ decay experiment.
The $\lambda_{4jk}^{\rm eff}$ can be constrained by various experiments such as $\mu \to e \gamma$, $\mu \to e e e$ and similar $\tau$ decays.
The bounds depend on the final lepton flavor indices ($j, k$), and currently the most severe bound comes from $\br (\mu \to eee) < 1.0 \times 10^{-12}$, which translates into $|\lambda^*_{i12} \lambda_{i11}| < (6.6 \times 10^{-7}) \times (m_{\til\nu_i} / 100 ~\gev)^2$ for $\til\nu_i$.
In a flavor-blind sense, it corresponds to $|\lambda^{\rm eff}_{4jk}| < 0.0008 \times (m_{\til\nu_4} / 100 ~\gev)$ for the $\til\nu_4$ LSP with the other $\til\nu$'s sufficiently heavy \cite{Barbier:2004ez}.
That is $|\lambda^{\rm eff}_{4jk}| < 0.0004$ ($0.0016$) for $m_{\til\nu_4} = 50 ~\gev$ ($200 ~\gev$), which falls into the ballpark of the aforementioned $\lambda^{\rm eff}_{4jk}$ value for $M/\left<S\right> = 1000$.
Larger $\lambda^{\rm eff}_{4jk}$ may be allowed for a specific choice of $(j, k)$ as the experimental bounds are flavor-dependent.
Because of many free parameters in $\Gamma_{\til\nu_4}$, a wide range of $\br(\til\nu_4 \to \ell^+_j \ell^-_k)$ for a given $\ell^+_j \ell^-_k$ can be accommodated.

\begin{figure*}[tb]
\begin{center}
\includegraphics[width=0.32\textwidth]{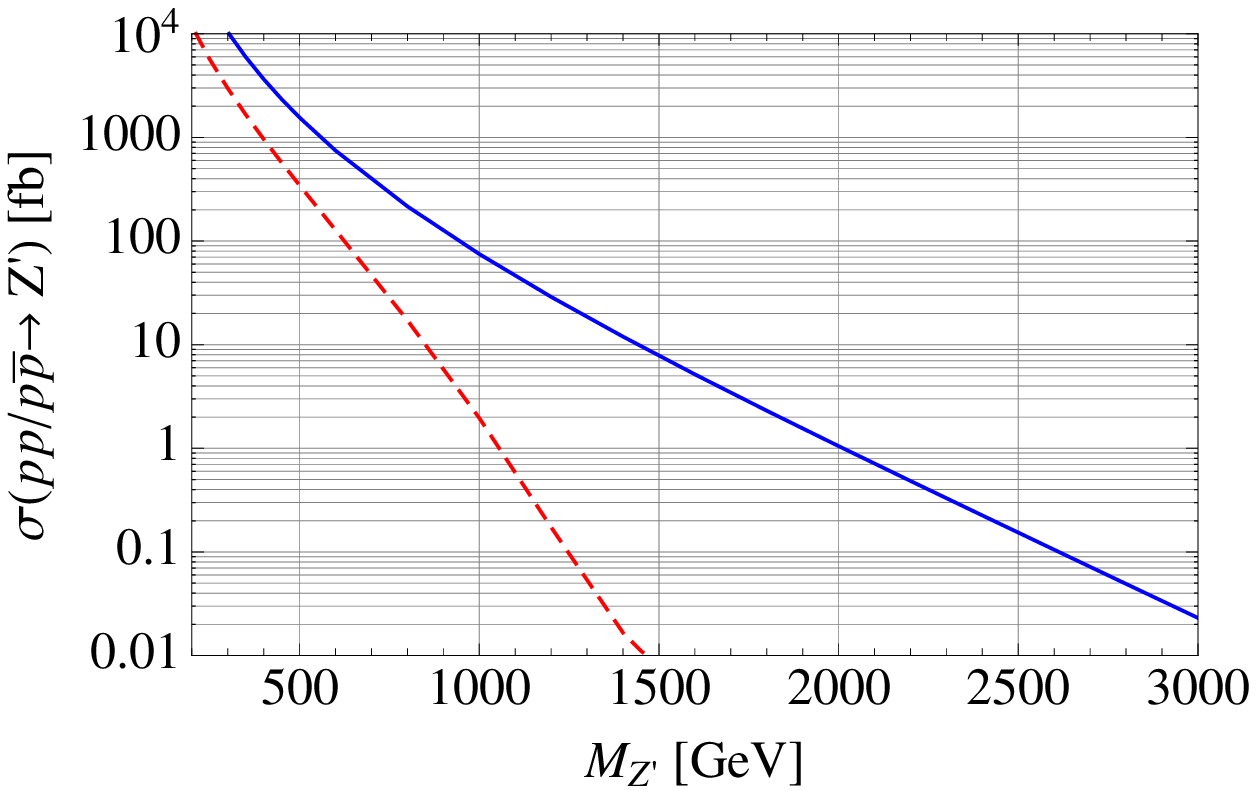} ~
\includegraphics[width=0.32\textwidth]{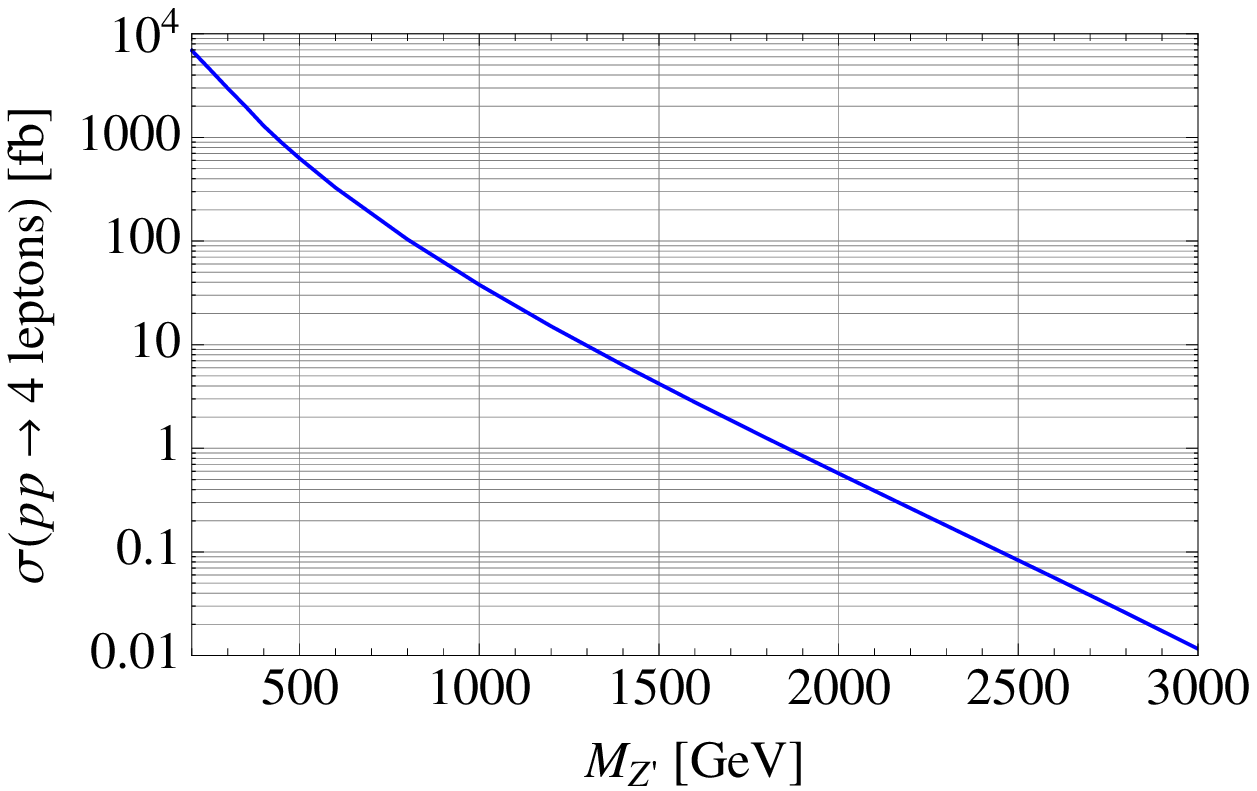} ~
\includegraphics[width=0.32\textwidth]{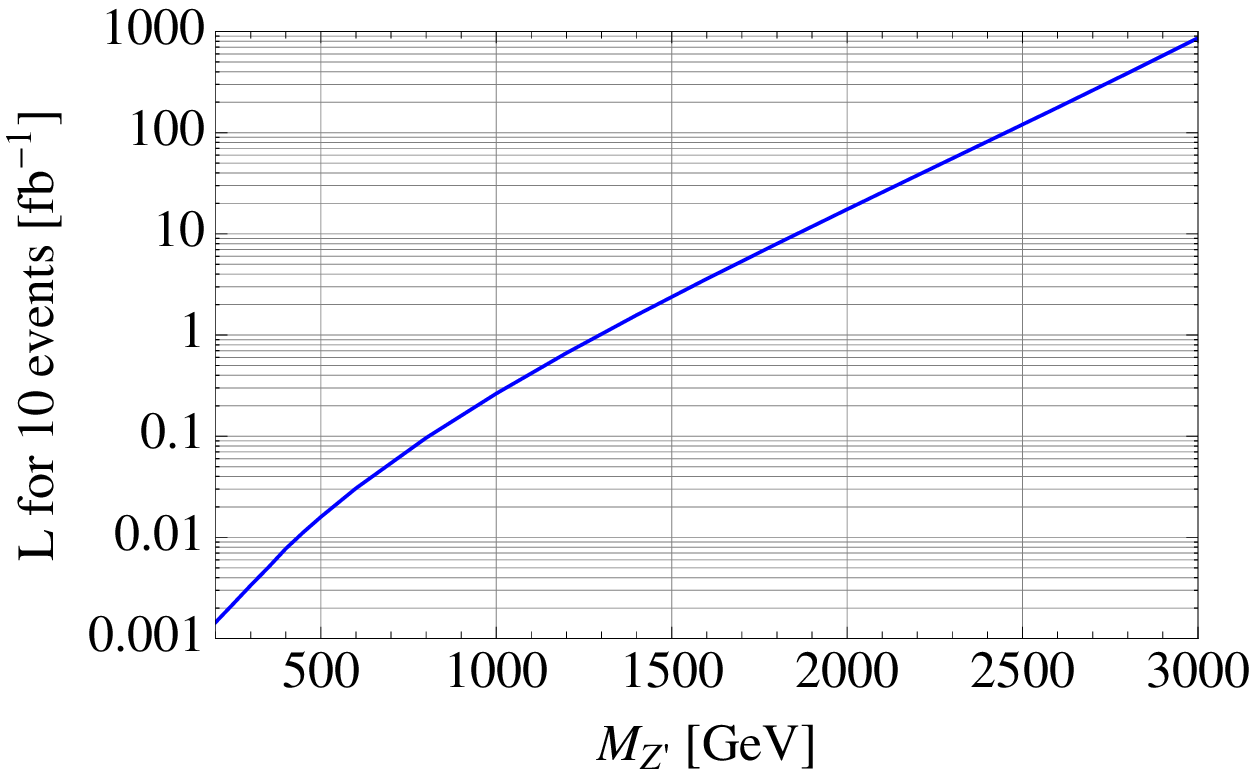} \\
(a) ~~~~~~~~~~~~~~~~~~~~~~~~~~~~~~~~~~~~~~~~~~~~~~~~~~ (b) ~~~~~~~~~~~~~~~~~~~~~~~~~~~~~~~~~~~~~~~~~~~~~~~~~~ (c)
\end{center}
\caption{(a) $Z'$ production cross section at the LHC7 ($E_\text{CM} = 7 ~\tev$) (solid curve) and the Tevatron ($E_\text{CM} = 1.96 ~\tev$) (dashed curve).
(b) Cross section of 4-leptons after cuts at the LHC7.
(c) Luminosity for 10 events after cuts at the LHC7.}
\label{fig:plots}
\end{figure*}

\section{COLLIDER PHENOMENOLOGY}
In this section we present quantitative cross section predictions of the 4-leptons channel for the LHC7 (LHC with $7 ~\tev$ CM energy) experiments.
For the calculations we use {\tt Comphep/Calchep} \cite{Pukhov:1999gg,Pukhov:2004ca}, with some modifications, and the parton distribution function of {\tt CTEQ6L} \cite{Pumplin:2002vw,modelfile}.

For definiteness, we take the $Z'$ gauge coupling constant to be $g_{Z'} = 0.1$; the $Z'$ production cross section and $Z'$ width can be simply scaled by $(g_{Z'}/0.1)^2$ for other $g_{Z'}$ values.
We assume that the $\til\nu_4$ LSP is the lightest 4G field, with $m_{\til\nu_4} = 50 ~\gev$, and that all new particles, except for the $\til\nu_4$ LSP, have masses larger than $M_{Z'} / 2$ so that $Z'$ decays only into the SM fermions and the $\til\nu_4$ pair.
Thus, the total $Z'$ width we take is the minimum value, which is $\Gamma_{Z'} \approx 1.6 \times 10^{-3} M_{Z'}$ for $M_{Z'} \gg m_{\til\nu_4}$.

The 4-lepton $Z'$ resonance cross section is
\beq
\sigma(p p  \to 4\ell) \simeq \sigma(p p \to Z') \br(Z' \to \til\nu_4 \til\nu_4^*) \br(\til\nu_4 \to 2\ell)^2 .
\eeq
The branching fraction is $\br(Z' \to \til\nu_4 \til\nu_4^*) \simeq 0.67$ for $M_{Z'} \gg m_{\til\nu_4}$.
The $\til\nu_4$ branching fractions to the light leptons ($ee$, $e\mu$, $\mu\mu$) are parameter dependent and flavor nonuniversality is expected.
We shall illustrate the case $\br(\til\nu_4 \to 2\ell) = 1$, which is indeed possible to arrange.

Figure~\ref{fig:plots} (a) shows the $Z'$ production cross section at the LHC (solid curve) and Tevatron (dashed curve), for the $Z'$ mass range $M_{Z'} = 200 - 3000 ~\gev$.
The low mass region would have been excluded by the dilepton $Z'$ resonance searches at the LHC had a $Z'$ coupled to the light leptons.
For instance, the current bound on the sequential $Z'$ model is already $M_{Z'} \gsim 1.8 - 1.9 ~\tev$ \cite{Collaboration:2011dca,CMS_dilepton} though its couplings are larger than our benchmark coupling.
The ratio of the Tevatron to LHC $Z'$ production cross sections is about $0.2$ for $M_{Z'} = 500 ~\gev$, and it drops rapidly at higher $M_Z'$.
Though it might be possible to have an observable 4-lepton resonance at the Tevatron, especially for the low $Z'$ mass region, we will focus on the LHC experiments in our analysis.

Figure~\ref{fig:plots} (b) shows the 4-lepton $Z'$ resonance cross section at the LHC after the following typical acceptance cuts and $Z'$ invariant mass cut: \\
$~~~~~~~$ (i) $p_T > 15 ~\gev$ ~(each lepton), \\
$~~~~~~~$ (ii) $|\eta| < 2.5$ ~(each lepton), \\
$~~~~~~~$ (iii) $|m_\text{inv}(4\ell) - M_{Z'}| < 3\Gamma_{Z'}$ ~(4-leptons).

The SM 4-lepton background to $ee$ and $\mu\mu$ pairs is principally from the $q \bar q \to Z Z$ subprocess.
As a recent ATLAS analysis shows, with nearly the same $p_T$ and $\eta$ cuts as ours, the SM background is negligible when the $m_{\til\nu_4}$ mass is outside the $Z$ window of $(66 - 116) ~\gev$ \cite{ATLAS_4leptons}.
Furthermore, some 4-lepton combinations (such as $eee\mu$, $e\mu\mu\mu$) do not have any significant SM backgrounds.
Thus, through all the $Z'$ mass range, we will require a small number of 4-lepton events (10 events) after the acceptance cuts, in order to estimate the discovery reach.

Figure~\ref{fig:plots} (c) shows the required luminosity at LHC7 to realize a signal of 10 events at a 4-lepton resonance as read from Fig.~\ref{fig:plots} (b).
For $g_{Z'} = 0.1$, an integrated luminosity at LHC7 of $L \simeq 17 ~\fb^{-1}$ is needed for discovery (10 events) of $M_{Z'} = 2 ~\tev$.
The existence of a 4-lepton $Z'$ resonance is already being probed at LHC7 in terms of the $M_{Z'}$ and $g_{Z'}$\footnote{A direct application of the ATLAS analysis $\sigma_{4\ell} < 4.7 ~\fb$ (with 3-events criteria and $L = 1.02 ~\fb^{-1}$) \cite{ATLAS_4leptons} gives $M_{Z'} \gsim 1500 ~\gev$ for $g_{Z'} = 0.1$.} and an integrated luminosity of $5 ~\fb^{-1}$ in each detector is expected before the end of 2011.
The current LHC dijet search results (with $L \sim 1 ~\fb^{-1}$) do not constrain the model for $g_{Z'} = 0.1 - 0.3$, as can be deduced from the estimates in Ref.~\cite{Lee:2011jk}.

A 4-lepton signal could be confused initially with a possible Higgs signal from $H \to ZZ$ with each $Z$ decaying to lepton pairs.
There are several distinguishing characteristics of the signals:
(i) $Z$ decay includes neutrino decay modes that are absent in $\til\nu_4$ decay;
(ii) $\til\nu_4$ can decay into different lepton flavors which allows final states like $eee\mu$ and $e\mu\mu\mu$, although these could be switched off by $\lambda_{412}^{\rm eff} = \lambda_{421}^{\rm eff} = 0$;
(iii) The angular distribution of leptons in their rest frame is flat for the scalars (Higgs and sneutrino), but $\theta$-dependent for the vectors ($Z$ and $Z'$);
(iv) If the $\til\nu_4$ mass differs from the $Z$ boson mass, the lepton pair invariant mass distributions from the sneutrino decays would peak at a value different from $M_Z$, either lower or higher; and
(v) $H \to ZZ$ should be accompanied by $H \to WW$, with a ratio of about 1 to 2.

Another exotic possibility for 4-lepton events is that a Higgs-like boson is produced via gluon-gluon fusion and it decays to a pair of hidden sector fields (vectors or scalars), each of which then decay to two leptons \cite{Gopalakrishna:2008dv,Falkowski:2010cm}.
The production cross section for a Higgs boson via gluon-gluon fusion would be much larger at LHC7 than at the Tevatron.

Though we have limited ourselves to only 4-lepton events, it is straightforward to extend the idea to other 4-fermion resonances depending on the values of $\lambda^{\rm eff}$ and $\lambda'^{\rm eff}$, such as $4\tau$, $2\ell + 2b$, $4t$, etc.

\section{SUMMARY}
We have discussed a novel $Z'$ search channel in which a 4-lepton $Z'$ resonance can be produced at the LHC without an accompanying 2-lepton $Z'$ resonance signal.
We have shown that it is possible to construct a consistent supersymmetric model which has a $Z'$ particle with this property.
The $U(1)$ symmetry of the model respects the baryon number for the first three generations.
The model is made anomaly free by the addition of a fourth generation of fermions.
Then the $Z'$ can decay to the fourth generation sneutrino pair, which in turn decay into lepton pairs, thus giving the 4-lepton resonance signal.
The $Z'$ and the $\til\nu_4$ can be discovered or excluded in the near future by the LHC experiments.

\acknowledgments
We thank the organizers, P. Fileviez-Perez and Y. Kamyshkov, of the Workshop on Baryon \& Lepton Number Violation (Gatlinburg, Tennessee, 2011 September), where some of this work was done. HL thanks S.-C. Hsu for useful discussions.
This work was supported in part by the U.S. Department of Energy under Grant Contract No.~DE-FG02-95ER40896 and No.~DE-AC02-98CH10886.


\end{document}